# A Novel Pre-processing Scheme to Improve the Prediction of Sand Fraction from Seismic Attributes using Neural Networks

Soumi Chaki, *Student Member, IEEE*, Aurobinda Routray, *Member, IEEE*, William K. Mohanty

*Abstract*— This paper presents a novel pre-processing scheme to improve the prediction of sand fraction from multiple seismic attributes such as seismic impedance, amplitude and frequency using machine learning and information filtering. The available well logs along with the 3-D seismic data have been used to benchmark the proposed pre-processing stage using a methodology which primarily consists of three steps: pre-processing, training and post-processing. An Artificial Neural Network (ANN) with conjugate-gradient learning algorithm has been used to model the sand fraction. The available sand fraction data from the high resolution well logs has far more information content than the low resolution seismic attributes. Therefore, regularization schemes based on Fourier Transform (FT), Wavelet Decomposition (WD) and Empirical Mode Decomposition (EMD) have been proposed to shape the high resolution sand fraction data for effective machine learning. The input data sets have been segregated into training, testing and validation sets. The test results are primarily used to check different network structures and activation function performances. Once the network passes the testing phase with an acceptable performance in terms of the selected evaluators, the validation phase follows. In the validation stage, the prediction model is tested against unseen data. The network yielding satisfactory performance in the validation stage is used to predict lithological properties from seismic attributes throughout a given volume. Finally, a post-processing scheme using 3-D spatial filtering is implemented for smoothing the sand fraction in the volume. Prediction of lithological properties using this framework is helpful for Reservoir Characterization.

## I. Introduction

HYDROCARBONS migrate from source rock through porous medium to reach reservoir rock for temporary preservation [1]. Finally, the mobile hydrocarbons get seized in the cap rocks. As such, the identification of hydrocarbon–enriched–formations by characterization of each layer in the borehole is of enormous importance to the explorers. Recognition of potential hydrocarbon–enriched zones in a prospective oil exploration field can be carried out using well logs which can categorize layers into different sections such as dry, water containing, and hydrocarbon bearing layers. The lithological properties in the neighborhood of a borehole can be estimated from well logs, whereas their distributions become difficult to predict away from the wells. In such cases, available seismic attributes can be used as a guidance to predict lithological information at all traces of the area of interest [2]. Well logs and seismic attributes are integrated at available well locations to design a reservoir model with the least uncertainty. However, mapping between well logs and seismic attributes is governed by nonlinear relationship and characterized by mismatch in information content. Such nonlinear problems can be approached using state-of-art computer–based methods like hybrid systems [3], [4], multiple regression, neural networks [5], Neuro-fuzzy Systems [6] etc.

It has been found from literature that ANN has been widely used by researchers and engineers from diverse backgrounds to model single or multiple target properties from predictor variables in different research problems because of its accurate prediction and generalization capability. For example, ANN has been used in climatological studies [7], ocean engineering [8], telecommunications [9], text recognition [10], financial time series [11], reservoir characterization [12]–[17], etc. A diverse dataset containing information assembled from multiple domains can be used for learning and validation of ANN. However, it has some inherent limitations. Firstly, performance of ANN is dependent on the selection of network structure and associated parameters. Secondly, training a complex multilayered network is a time intensive process. Furthermore, a complex network trained with relatively smaller number of learning patterns may lead to over fitting. It is equally important to assess the possibility of modelling the target property from predictor variables using ANN or any other nonlinear modelling approach. Sometimes the model performance can be improved by applying suitable filtering techniques to the predictor/target variables in the pre-processing stage. Several studies have contributed to the performance analysis of ANN along with other machine learning algorithms to model a target variable from single or multiple predictors with respect to RC

Soumi Chaki and Aurobinda Routray are with the Department of Electrical Engineering, Indian Institute of Technology Kharagpur, Kharagpur- 721302, India (e-mail: soumibesu2008@gmail.com, aroutray@ee.iitkgp.ernet.in).

William K. Mohanty is with the Department of Geology and Geoscience, Indian Institute of Technology Kharagpur, Kharagpur- 721302, India (e-mail: wkmohanty@gg.iitkgp.ernet.in).

problem; however, the following aspects still remain unexplored:
- Design of an appropriate pre-processing stage for effectiveness of machine learning algorithms
- Proper choice of structure and parameters associated with selected machine learning algorithms (here, ANN model parameters- e.g. activation function type, number of hidden layers etc.)
- Suitable post-processing methods for the predicted output

This paper proposes a novel pre-processing scheme and demonstrates the use of the said scheme in an appropriately designed framework, consisting of three stages- pre-processing, modelling, and post-processing, to estimate sand fraction (SF) from multiple seismic attributes. Here the target variable SF represents per unit sand volume within the rock.

The workflow starts with pre-processing stage, which uses three alternative approaches for target variable regularization based on Fourier Transform (FT), Wavelet Decomposition (WD), and Empirical Mode Decomposition (EMD). Next, a functional relationship between the regularized target sand fraction and the seismic attributes is calibrated using ANN. In this study, a simple network structure consisting a single hidden layer is selected over relatively complex networks. The network parameters and training algorithm are decided empirically. The results obtained, are evaluated in terms of four performance indicators– Correlation Coefficient (CC), Root Mean Square Error (RMSE), Absolute Error Mean (AEM) and Scatter Index (SI) [7]. The inclusion of the pre-processing stage in the workflow is found to produce remarkable improvement in the prediction results as opposed to use of the original sand fraction using the same ANN structure. After learning and validation, post processing is carried out to improve the prediction result over the study area. In this way, a complete workflow is devised for modelling a lithological property (here, sand fraction) from given seismic attributes.

The paper is organized as follows. Section II describes the study area and the complete workflow adopted in this paper. Section III explains the pre-processing steps. Section IV illustrates the ANN model building and validation. In Section V, the post-processing is described. Finally, Section VI concludes the paper with a discussion on achievements in this study and possible future directions of research.

## II. STUDY AREA AND WORKFLOW

### A. Study Area

In this study, the working dataset has been acquired from a western onshore hydrocarbon field in India. The hydrocarbon field is located at an intra-cratonic basin, spread along the western periphery of central India. It is surrounded by the Aravalli range. Deccan craton and Saurashtra craton are located in the east-west direction along the basin. Structurally, the field is located as a broad nosing feature; thus, housing the hydrocarbons between two major synclines. The hydrocarbon is present within a series of vertically stacked sandstone reservoirs individually separated by intervening shale. The average thickness of a sand layer is in the order of 5-6 m. Due to the discrete sand depositions in thin layers, imaging of the seismic data and obtaining a mapping between seismic and borehole dataset are challenging tasks.

A spatial database containing seismic attributes and well logs has been acquired from the study area in .sgy and .las format respectively. The SEG-Y file format is one of the numerous standards developed by the Society of Exploration Geophysicists (SEG) for storing geophysical data [18]. The Log ASCII Standard (LAS) format has been developed by the Canadian Well Logging Society to standardize the organization of digital log curves information in 1989 [19]. As the workflow is developed on MATLAB platform, the .sgy data files are converted in .mat format (MATLAB platform compatible format) for MATLAB compatibility.

Fig. 1 represents the placements of the four wells across the study area. These wells will be hereafter referred as A, B, C, and D in terms of inlines and crosslines (xlines). Additionally, well A (XXX3645, XXX7387), well B (XXX2107, XXX7285), well C (XXX2768, XXX7073), and well D (XXX0916, XXX7807) are the locations in terms of Universal Transverse Mercator (UTM) coordinates. The depth of the each well is around 3000 meter from ground, whereas the zone of interest is from 2750 meter to 2975 meter in subsurface for well A. The zone of interests are 2720 meter – 2950 meter for well B, C, and D.

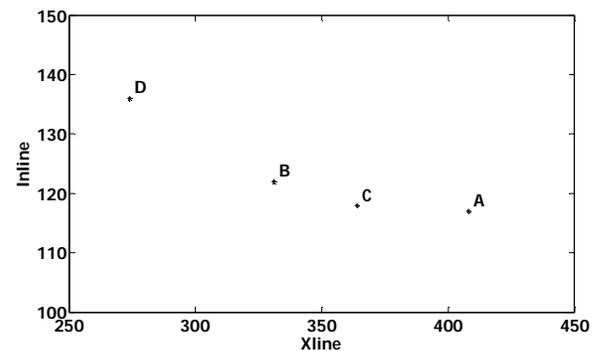

Fig. 1. Location of four wells in the study area in terms of inlines and crosslines (xlines)

The borehole dataset contains basic logs such as gamma ray, resistivity, density along with derived geo-scientific logs such as sand fraction, permeability, porosity, water saturation, etc. These well logs are treated as one dimensional signals for further processing in this study. On the other hand, the seismic dataset contains seismic impedance, amplitude, and instantaneous frequency throughout the volume.

### B. Workflow

The systematic workflow consisting three stages e.g. pre-processing, model building–validation and post-processing is demonstrated in Fig. 2. The workflow is implemented on a 64-bit MATLAB platform installed in Dell Precision M6800 with Intel® core™ i7-4700MQ CPU @2.40 GHz having 32GB RAM. The regularization approach in the pre-processing stage is highlighted using a small dotted rectangle in Fig. 2.

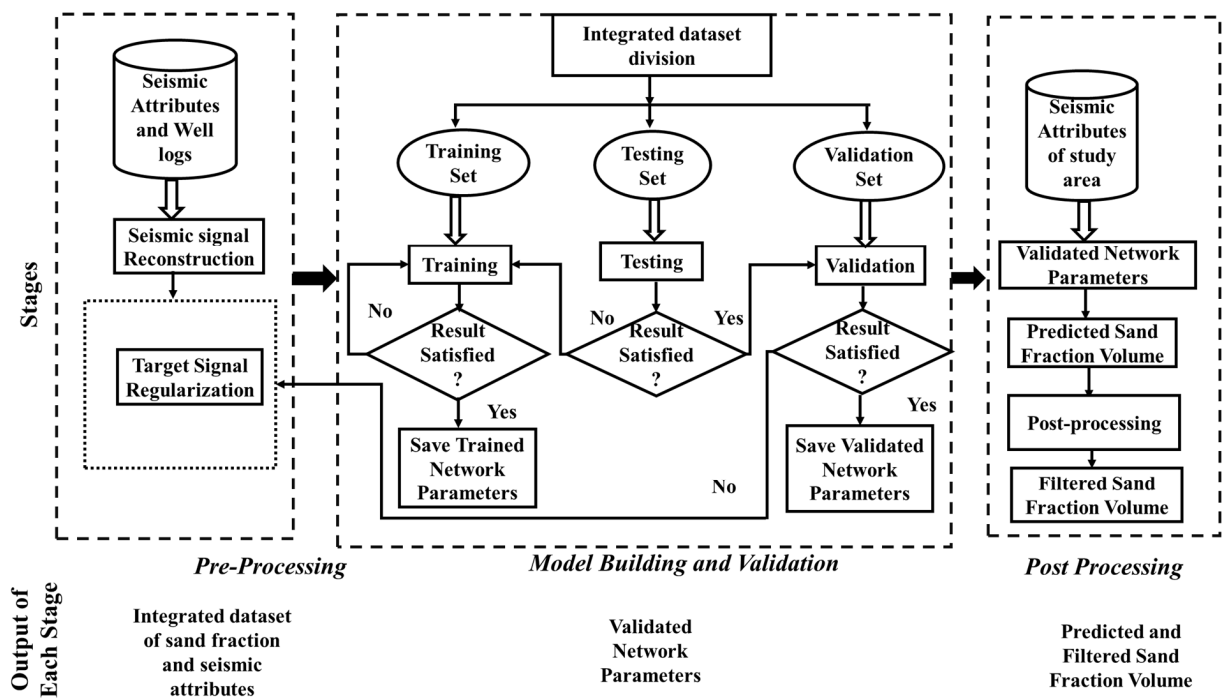

Fig. 2. The proposed workflow to model sand fraction from multiple seismic attributes

Three alternative regularization methods based on FT, WD, and EMD are implemented in this study. In the model building stage, training, testing, and validation are carried out consecutively. If the validation performance does not improve as expected even after implementing either of the three regularization techniques, then the parameters associated with the selected regularization method are modified before carrying out training, testing, and validation again. This tuning is carried out using a trade-off between loss of high frequency information from target attribute and acceptable performance evaluators.

The improvement in the ANN performance with the regularization technique is quantified using the performance evaluators presented in Tables-III and IV later. First, the ANN is trained with the original sand fraction as target variable without involving any pre–processing. The same procedure is repeated with the regularized sand fraction as target variable of the ANN. This is helpful to reassure the improvement in ANN performance while using the regularized target variables. The validated network is used to obtain sand fraction over the study area from the three seismic attributes. The post-processing stage includes further smoothing and filtering of the estimated output to remove unwanted predictions.

## III. Pre-processing

Pre-processing plays a crucial role on the performance of a machine learning algorithm. In this paper, an efficient pre-processing approach is proposed as part of the adopted methodology to obtain a functional relationship between seismic attributes and lithological properties.

### A. Signal Reconstruction

The borehole data are recorded at specific well locations along depth with a high vertical resolution. Conversely, the seismic data are collected spatially in time domain with a sampling interval of two milliseconds. Signal reconstruction is an important pre-processing step and has been discussed in existing literature. In [20], the up-scaling of an acoustic impedance log has been carried out by a wavelet transform based method as well as a geostatistical technique. The acoustic impedance log is computed from the recorded logs of the P-wave velocity and the density at the corresponding well-location. The reconstructed acoustic impedance logs obtained from the raw impedance log by wavelet transform and geostatistical methods have turned out to be similar. In [21], the seismic attributes and well logs have been integrated to predict facies using a data-driven methodology. The complexities regarding resolution of individual datasets (seismic and well logs) and difference in the individual sampling intervals have been discussed in detail in [21]. For the present work, the well logs are converted from depth to time domain at 0.15 milliseconds sampling interval using the given velocity profile obtained from well-seismic-tie. It can be noted that the sampling intervals of the two datasets (seismic attributes and well logs) are different. Hence in order to integrate the two, the band limited seismic attributes are reconstructed at each time instant corresponding to the well logs by a sinc interpolator while adhering to the Nyquist–Shannon sampling theorem [22].

Fig. 3 shows the three seismic attributes and the sand fraction, along well A. The red dots on the seismic attributes represent their original values at time interval of two milliseconds and the green curves represent the corresponding reconstructed signals at the time instants marked on the well logs. The blue high frequency curve in Fig. 3(d) represents sand fraction along the same well.

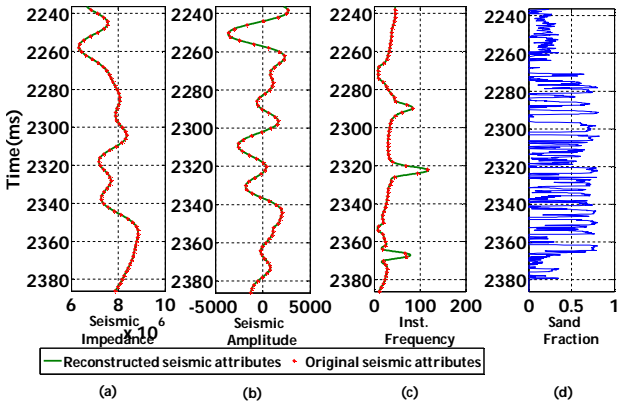

Fig. 3. Seismic attributes and sand fraction along well A (The red dots on the seismic attributes are the original values, the green line is the reconstructed signal)

## B. Signal Regularization

It can again be observed from Fig. 3 that the frequencies present in seismic signals are much lower compared to that of the sand fraction. In other words, the sand fraction carries much more information as compared to the seismic attributes. According to laws of information theory, a higher information-carrying signal cannot be modelled using single or multiple lower information-carrying predictor signals [23]. Only a part of the target variable that is dependent on the predictor variables can be modelled. For the present case, the spatially distributed seismic attributes have low vertical resolution; whereas the high resolution borehole data are recorded at the specific well locations and carry high information content [24]. Hence, the estimation of subsurface lithological properties from seismic attributes deals with uncertainty due to the mismatch in the information content. In order to circumvent this disparity, the seismic attributes are re-sampled at each time instants corresponding to the well logs to integrate with the lithological properties. However, this re-sampling does not contribute any additional information [23]. The seismic information would not be able to successfully delineate the rock properties completely, which is a key step for RC. As no new information can be added to the seismic attributes, some amount of information has to be filtered from the high information carrying SF signal in order to calibrate a functional relationship between them. Thus, the necessity of information filtering for the SF prediction through regularization is established.

In this paper, three different signal processing approaches are selected and implemented in order to filter the target signal. The parameters belonging to this stage are tuned following the changes in entropy before and after filtering along with visual inspection of the output signal with respect to that of the original target signal. The entropy has been computed from the Power Spectral Density (PSD) of the signal. The average amount of information gained from a measurement that specifies $X$ is defined to be the entropy $H(X)$ of a system. It can be formally defined as

$$H(X) = -\sum_i p(x_i) \log_2 p(x_i) \quad (1)$$

where $p(x_i)$ is the probability of $X$ having the $i^{th}$ value $x_i$ in the dataset. This is known as Shannon entropy [23], [25]. If $A$ is another random variable described on the same dataset then the mutual information between the two can be expressed as

$$I(X;A) = H(X) - H(X|A) \quad (2)$$

where, $H(X|A)$ is the conditional entropy of $X$ after $A$ has been observed. A reservoir property (SF for the present case) can be represented by $X$ and seismic attribute e.g. seismic impedance can be represented by $A$. The statistical property of $I(X;A)$ can be interpreted as the reduction in the uncertainty of the reservoir property, due to observation of attribute $A$. In [26], Normalized Mutual Information (NMI) is defined as the mutual information normalized by minimum entropy of both the variables.

$$NMI(X;A) = I(X;A) / \min(H(X), H(A)) \quad (3)$$

In this study, the NMI computed between predictor and target signal has been used to adjust the parameters of the information filtering algorithms.

Before starting the regularization, the seismic inputs and sand fraction values are normalized by Z-score and min-max methods respectively using the data from all four wells taken together. In this paper, the modelling of the target SF from the three seismic attributes is carried out by a single hidden layer ANN trained using a back propagation algorithm. The network structure and training procedure are discussed in detail in the following section. The target variable is normalized within the range of output activation function keeping some offset from limiting value of the activation function. Otherwise, the back propagation algorithm tends to drive the free network parameters to infinity. As a result, the learning process may slow down [27]. Hence, the target variables are normalized between 0.1 and 0.9 so that it does not overlap with the saturation region of the log-sigmoid function.

### 1) Fourier Transform (FT)

The first regularization approach is based on Fourier transform (Algorithm I). Here, the spectrums of target and predictor variables are compared and higher frequency components of the target signal are truncated. Then, the target signal is reconstructed using Inverse Fourier transform (IFT).

---

**ALGORITHM I**: SF regularization Based on Fourier Transform

**Task**: Regularizing target sand fraction based on Fourier Transform

**Input**: Predictor signal $x(t)$ and target signal $y(t)$

a) The sand fraction $y(t)$ extracted from a well-log by proper depth to time conversion using the given velocity profile. Let the seismic amplitude along the same well-log be $x(t)$.

b) Compute Fourier Transform of $x(t)$: $X(k) = \sum_{j=1}^{N} x(j)\omega_N^{(j-1)(k-1)}$

where, $\omega_N = e^{-2\pi i/N}$ is an $N^{th}$ root of unity

Similarly, FT of target $y(t)$ is computed as: $Y(k) = \sum_{j=1}^{N} y(j)\omega_N^{(j-1)(k-1)}$

where, $\omega_N = e^{-2\pi i/N}$ is an $N^{th}$ root of unity

c) Compare the spectrums of target and predictor signals

d) Select the bandwidth parameter $\zeta_{max}$ Hz

---

e) The part of the target spectrum exceeding $\xi_{max}$ Hz is truncated to zero.
   Modified Target : $Y_{mod}(k)$

f) Construct regularized target signal $y_r(t)$ by carrying out IFT of the truncated spectrum : $y_r(t) = \frac{1}{N}\sum_{k=1}^{N} Y_{mod}(k)\omega_N^{-(j-1)(k-1)}$, where, $\omega_N = e^{-2\pi i/N}$ is an $N^{th}$ root of unity

g) Calculate entropies of predictors (seismic attributes here) as well as original and regularized target signals (sand fraction here).

h) If entropy of regularized target is comparable with that of the predictor signal and the regularization result is satisfactory, then regularization is completed else go to step d).

**Output :** Regularized target signal $y_r(t)$

Comparison between the spectrums of the sand fraction (Fig. 4 (a)) and that of the seismic impedance (Fig. 4 (b)), reveals the presence of higher order frequencies in the former.

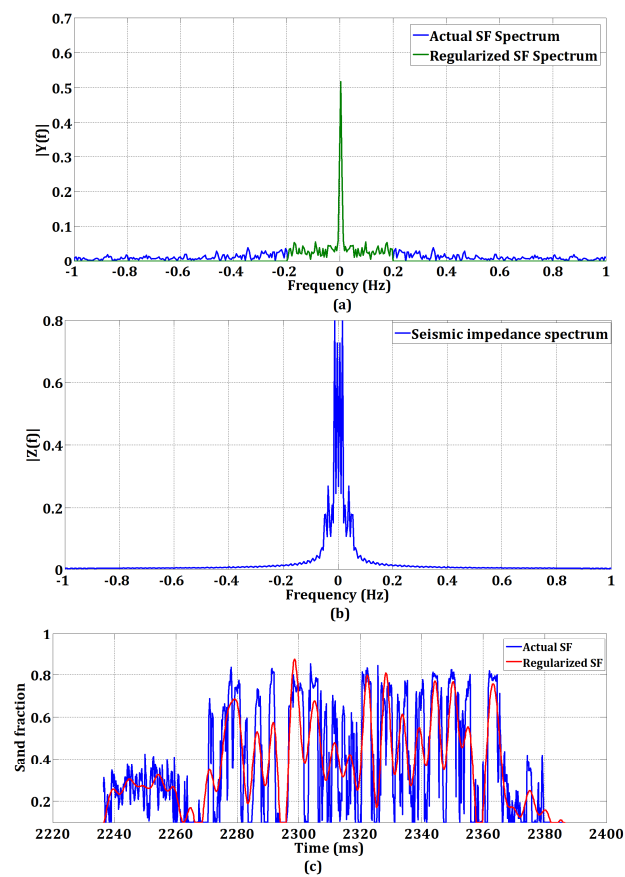

Fig. 4. Regularization based on FT to reconstruct target signal along Well A; (a) SF spectrum, (b) seismic impedance spectrum, and (c) superimposed plot of actual and regularized SF

It can be observed from Fig. 4(b) that the spectrum of band-limited seismic impedance diminishes beyond a certain frequency range (-0.2: +0.2 Hz). Then, the part of the sand fraction spectrum belonging to a slightly wider frequency range (green curve, Fig. 4(a)) is reconstructed to obtain the regularized target. The wider range of frequency is chosen with the assumption that ANN as a nonlinear predictor is capable of mapping input signals of lower frequencies to output signals with higher frequencies. Of course it needs an entirely different research to find the prediction capability of a given nonlinear mapping process. The original and regularized SF signals are presented in Fig. 4(c) by the blue and red curves respectively.

The FT based regularization method is more sophisticated than bandpass filtering in the time domain. In order to carry out bandpass filtering, two cut-off frequencies of the ideal bandpass filter are defined beyond which all frequency components are removed. In case of ideal bandpass filters, zero phase shift is an important requirement. However, ideal bandpass filter is not realizable while working with practical datasets having finite number of samples. Moreover, in case of practical time-domain bandpass filtering, a transition takes place between the pass band and the stop band i.e. the frequency component beyond the cut-off frequencies does not diminish abruptly. In case of FT based regularization, the frequency components of the SF spectrum beyond the selected parameter range (-0.2:+0.2 Hz) are completely truncated before reconstruction of regularized SF log. Additionally, the phase shift resulting from bandpass filtering can yield lead-lag relationships between actual and filtered logs [28]. On the other hand, FT based regularization does not incur phase shift between actual and regularized logs as shown in Fig. 4(c).

TABLE I
ENTROPY (IN BIT) OF PSD OF SIGNALS FOR WELL A

| Variables | | Entropy Value |
|---|---|---|
| Seismic Impedance | | 0.15 |
| Seismic Amplitude | | 0.16 |
| Inst. frequency | | 0.12 |
| Original SF | | 0.28 |
| Regularized SF | FT | 0.17 |
| | WD | 0.20 |
| | EMD | 0.24 |

TABLE II
NORMALIZED MUTUAL INFORMATION (NMI) AMONG PREDICTORS AND TARGET SAND FRACTION FOR WELL A

| Predictor Variable | | Seismic Impedance | Seismic Amplitude | Inst. Frequency |
|---|---|---|---|---|
| Original Sand Fraction | | 0.12 | 0.11 | 0.09 |
| Regularized SF | FT | 0.16 | 0.16 | 0.14 |
| | WD | 0.15 | 0.14 | 0.11 |
| | EMD | 0.14 | 0.14 | 0.12 |

As shown in Table I, the information content of the original sand fraction is higher as compared to that of the seismic predictor variables which makes it difficult to model the target (sand fraction) from predictor attributes. The regularization process decreases the information content in the sand fraction as seen in Table I. The dependency between predictor and target variables in terms of NMI (Table II) also improves as a result of regularization.

*2) Wavelet Decomposition (WD)*

Wavelet decomposition has been extensively used in different fields of research. For example, it is applied in EEG signal for artifact removal [29], [30] and study of geomagnetic signals [31]–[33] etc. A time-frequency representation of a one dimensional non-stationary signal is obtained using wavelet analysis. Wavelets are oscillating functions localized in time and frequency [29], [34], [35]. A finite energy time domain signal can be decomposed and expressed in terms of scaled and shifted versions of a mother wavelet $\psi(t)$ and a corresponding scaling function $\phi(t)$ after decomposition. The scaled and shifted form of the mother wavelet $\psi_{l,k}(t)$ and the corresponding scaling function $\phi_{l,k}(t)$ are mathematically represented as

$$\psi_{l,k}(t) = 2^{l/2}\psi(2^l t - k), l,k \in R \quad (4)$$

$$\phi_{l,k}(t) = 2^{l/2}\phi(2^l t - k), l,k \in R \quad (5)$$

The original signal $X(t)$ is first decomposed into high frequency and low frequency components using high pass and low pass filters. After each filtering step, the output time series is down-sampled by two. The low frequency part approximates the signal while the high frequency part denotes residuals between original and approximate signal. At successive levels the approximate component is further decomposed using the same set of high-pass and low-pass filters. A time domain signal $X(t)$ can be expressed in terms of the aforementioned mother wavelet $\psi_{l,k}(t)$ and corresponding scaling function $\phi_{l,k}(t)$ at level $l$ as

$$X(t) = \sum_k a_l(k)\phi_{l,k}(t) + \sum_k d_l(k)\psi_{l,k}(t) \quad (6)$$

where $a_l(k)$ and $d_l(k)$ are the approximate and detailed coefficients at level $l$. These coefficients are computed using filter bank approach as in [36]. Fig. 5 describes the steps of wavelet decomposition for three levels. Here, a signal is decomposed into approximate and detailed coefficients using low pass $H(k)$ and high pass $G(k)$ filters respectively. After decomposition, the coefficients can be modified. In case of signal reconstruction, the modified approximate and detailed coefficients are up sampled by two and then convolved with respective synthesis filters and then the resulting pair is summed. Finally, modified signal is acquired following $l$ level synthesis.

The performance of wavelet analysis is dependent on the mother wavelet selection and decomposition level. The Daubechies family of wavelets has a compact support with relatively more number of vanishing moments [34]. Therefore, in most of the cases different variants of Daubechies family wavelets are used for signal analysis. The initial choice for mother wavelet and decomposition level can be db4 and six respectively for this kind of study. However, the initial choice of wavelet and decomposition level can be modified if the regularization results are not satisfactory. In this paper, the choice of mother wavelet and decomposition level has been carried out empirically based on the regularization result and performance of the ANN. Algorithm II describes the steps associated with WD based regularization.

Fig. 6(a)-(b) represent the results of WD–based regularization of target sand fraction with predefined wavelet type, and decomposition level. The first three detailed coefficients of original sand fraction signal are demonstrated in Fig. 6 (a). After the decomposition, detailed coefficients of initial levels of original target signal are made zero and regularized signal is constructed by performing Inverse Discrete Wavelet Transform (IDWT) from the modified coefficients.

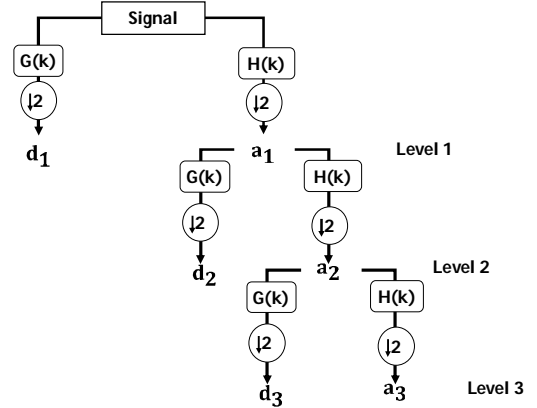

Fig. 5. Demonstration of wavelet decomposition of a signal for level 3

---

**Algorithm II**: SF Regularization Based on WD
**Task** : Regularizing target SF based on Wavelet Decomposition
**Input** : Predictor signal $x(t)$ and target signal $y(t)$
  a) Same as Algorithm-I.
  b) Select the wavelet type and number of decomposition levels.
  c) Apply the procedure as in Fig. 5 to the target signal.
  d) Decide: detailed coefficients to be truncated for regularization
  e) The selected detailed coefficients are made zero.
  f) The regularized target signal is reconstructed from the modified coefficients.
  g) Calculate entropies of predictors as well as original and regularized target signals.
  h) If entropy of regularized target is comparable with that of the predictor signal and the regularization result is satisfactory, then regularization is completed, else go to step d).

**Output :** Regularized target signal $y_r(t)$

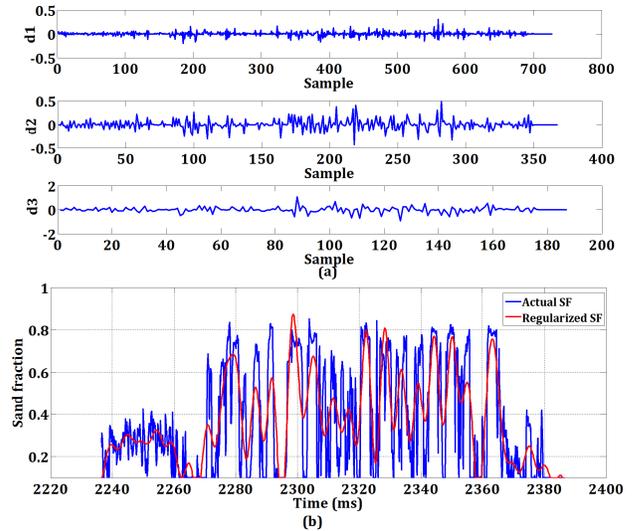

Fig. 6. Regularization based on WD to reconstruct the sand fraction along Well A; (a) decomposition result of the actual SF (d1-d3), (b) superimposed plot of actual and regularized SF

For Well A, first five detailed coefficients are truncated and regularized target is reconstructed from approximate and detailed coefficients of the sixth level by IDWT. The regularization result for Well A is presented in Fig. 6 (b), where blue and red curves represent original and regularized target sand fraction signals respectively. Table I-II reveal the changes in information content of the original and regularized sand fraction by WD and increase of dependency between target and predictor variables as a result of regularization.

*3) Empirical Mode Decomposition (EMD)*

Seismic and well log signals are non-stationary signals. Reports suggest that in most of the cases the frequency analysis of signals are carried out in selected windows with respect to a given orthogonal basis [37]–[39]. The disadvantage of basis decomposition techniques is the mismatch between signal trend and constant basis functions. These necessitate a new decomposition method, namely Empirical Mode Decomposition (EMD). EMD is an algorithmic decomposition method which decomposes the input signal into a set of Intrinsic Mode Functions (IMFs) and a residue signal [40]. There are two properties associated with IMFs such that (1) the numbers of zero–crossings and extrema present in IMFs are same, and (2) IMFs are symmetric with respect to the local mean [40]. In other words, EMD detects and extracts the highest frequency component in the signal [34], [41]–[47]; such that, in step ($k$+1), the extracted IMF contains lower frequency component compared to that extracted in step $k$. Moreover, being an adaptive data-driven method, EMD decomposes an input signal into a variable number of components. Thus, EMD overcomes the inherent limitation of deciding *a priori* the number of decomposition levels as in WD. Algorithm III describes the detailed steps associated with EMD based regularization of target SF.

---

**Algorithm III** SF Regularization Based on EMD

**Task** : Regularizing target sand fraction based on EMD
**Input** : Predictor signal $x(t)$ and target signal $y(t)$

a) Same as Algorithm I.
b) Initialize: $r_0(t) = x(t)$, $i = 1$
c) Extract the $i^{th}$ IMF:
   i. Initialize: $h_0(t) = r_i(t)$, $j = 1$
   ii. Extract the local minima and maxima of $h_{j-1}(t)$
   iii. Create upper envelope $e_{\max}(t)$ and lower envelope $e_{\min}(t)$ of $h_{j-1}(t)$ by interpolating local maxima and minima
   iv. Calculate mean envelope: $m_{j-1}(t) = \dfrac{e_{\max}(t) + e_{\min}(t)}{2}$
   v. $h_j(t) = h_{j-1}(t) - m_{j-1}(t)$
   vi. If $h_j(t)$ is an IMF, then, $imf_i(t) = h_j(t)$
       else go to step- (ii). with $j = j + 1$
d) $r_i(t) = r_{i-1}(t) - imf_i(t)$
e) If $r_i(t)$ has at least two extrema, then, go to c) with $i = i + 1$
   else, $x(t) = \sum_{i=1}^{n} imf_i(t) + r_n(t)$ is decomposed into $n$ numbers of IMFs and residue signal.
f) EMD of target signal $y(t)$ is carried out following steps b)-e)

$$y(t) = \sum_{i=1}^{p} imf_i(t) + r_p(t)$$

g) The number of IMFs and distribution of IMFs are observed for target $y(t)$ and predictor $x(t)$: $p > n$
h) Decide $p_1$: number of IMFs truncated from the EMD of $y(t)$ for regularization where, $p_1 < p$
i) Construct regularized target signal $y_r(t)$:

$$y_r(t) = \sum_{i=1}^{p_1} imf_i(t) + r_p(t)$$

j) Calculate entropies of predictors as well as original and regularized target signals.
k) If entropy of regularized target is comparable with that of the predictor signal and the regularization result is satisfactory, then regularization is completed else go to step h).

**Output** : Regularized target signal $y_r(t)$

---

In Fig. 7 (a)-(b), the first four IMFs of the sand fraction log and the IMFs and residue of the decomposed seismic impedance along the Well A are plotted.

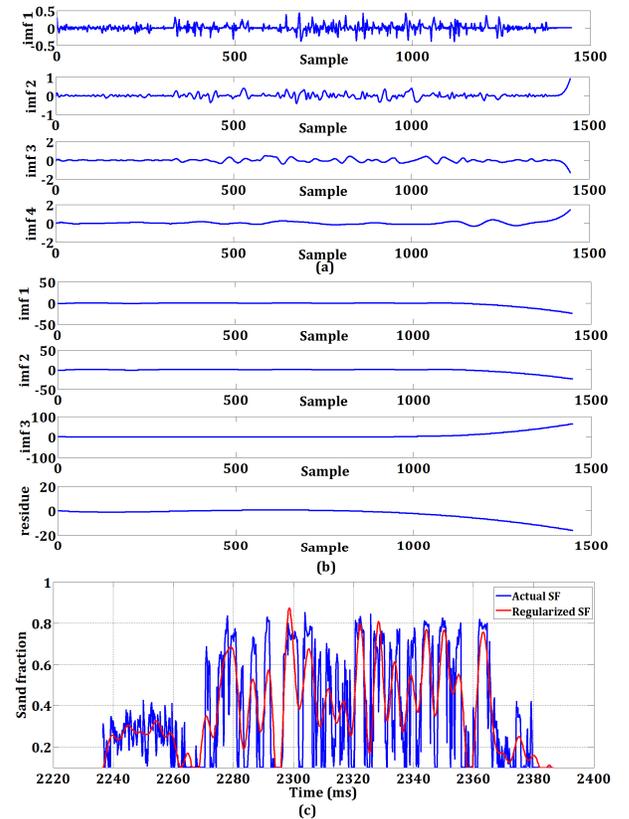

Fig. 7. EMD–based regularization to reconstruct sand fraction signal along Well A; (a) IMF1-IMF4 of the decomposed SF, (b) IMF1-residue of the decomposed seismic impedance (c) superimposed plot of actual and regularized SF

The comparison between Fig. 7 (a)-(b) reveals that the number of IMFs obtained is higher in case of target signal (sand fraction) than its predictor counterpart (seismic-impedance). The first IMF component is suppressed and the other IMFs are used to reconstruct the regularized sand fraction. The superimposed plots of actual (blue curve) and regularized sand fraction (red curve) signals are presented in Fig. 7 (c). The user decides the regularization result is satisfactory or not based on visual inspection of original and regularized target variables. The regularized target is smoother compared to original signal;

nevertheless, the trend of the original signal is preserved even after information filtering based on either of the three proposed regularization methods. The number $p_1$ is selected such that it is less than that of the decomposed IMFs of the original signal to be regularized. It is usually selected as one less than the number of decomposed IMFs of the original signal. If the regularization result is satisfactory in terms of entropy criterion and superimposed plots of actual and regularized SF, the initial $p_1$ value is retained as in Algorithm III. Otherwise, the value of $p_1$ is to be modified. The initial selection of $p_1$ would be one less than the number of decomposed IMFs of the original target signal for the working datasets. Table I represents the entropies of predictor and target attributes for Well A. The improvement in mutual dependency between predictor and target variables in terms of NMI is evident from Table II.

The FT is a linear transform where the frequency spectrum of a time-domain signal is obtained. Based on the comparison between the frequency spectrums of the predictor seismic impedance and target SF, the regularization parameter is selected. The result of FT based regularization changes according to the value of the selected regularization parameter. If the value of the selected regularization parameter were very small, then the regularized signal would become smooth by shedding the finer trends of the original target signal. Again, regularization with large parameter value would not carry out sufficient information filtering to enable effective modelling of the target variable. The basic assumption in FT is stationarity of a signal. However, seismic and SF signals are non-stationary. Therefore, the next two approaches based on WD and EMD are opted to cross-validate the performance of FT based regularization. The selection of the mother wavelet and the decomposition level determines the granularity of the decomposed target signal. The regularized target signal is reconstructed from the modified coefficients. These parameters (wavelet type, decomposition level, and detailed coefficients to be truncated) are finalized empirically based on the regularization result. However, in case of EMD, the number of parameters is less. Only, the number of IMFs to be used for reconstruction of the target property needs to be decided. In all the three cases, it is ensured that the regularized target signal retains the trend of the original signal. In case of FT based approach, the regularization parameter is decided based on the FT results of both the predictor as well as the target signal. The number of IMFs of the seismic attributes is less compared to those of SF in case of EMD. While in both the cases of WD and EMD based regularization, the number of detailed coefficients and IMFs are decided irrespective of the decomposition results of the predictor seismic signals. Instead these parameters are decided based on the regularization results and entropy criterion as mentioned in Algorithm II and Algorithm III respectively.

## IV. Model Building and Validation

In order to establish the efficacy of the proposed regularization method, the task of model building and validation stage is carried out using both the original and regularized sand fraction as target variable. The training dataset is created by aggregating 70% sample patterns from each of the wells. The training patterns are scrambled to remove any possible trend along the depths. The remaining 30% samples from each of the four wells are combined and then divided into two parts to create the testing and the validation datasets. First, the network is trained using training patterns with initial parameter values. Then, the network structure and activation functions are tuned using testing patterns. The testing phase is important for evaluating the generalization capability of the trained network [11]. The network which performs satisfactorily in terms of CC, RMSE, AEM and SI is then selected for use in the validation stage. The statistical analysis of the errors involved in the model are important for proper understanding of the performance. The initial network structure is decided intuitively depending on nature of the problem and amount of available training patterns. In this study several runs of training, testing and validation of neural network structures with varying number of neurons, layers, activation functions and learning methods have been carried out to decide the best structure as well as the most effective learning algorithm. Finally in the hidden layer, hyperbolic tangent sigmoid transfer function has been used. The tangent sigmoid transfer function is an automatic choice for researchers for use in hidden layer to achieve the bi-directional swing [48], [49]. The activation function used in the output layer is log-sigmoid which is non-symmetric in nature and facilitates faster learning rate [27]. The number of nodes in the input layer is same as the number of predictor attributes to be used to model the ANN. For example, in case of predicting sand fraction from three predictor attributes– namely, seismic impendence, instantaneous frequency and amplitude, the number of input and output nodes will be three and one respectively. Finally, a network with single hidden layer is trained using the Scaled Conjugate Gradient (SCG) Back propagation Algorithm [50]. The back-propagation learning using the SCG method is given in the Appendix. SCG proceeds in the conjugate direction of the previous step instead of following the gradient direction as in other gradient descent based algorithms. In case of the conjugate gradient algorithms, the step size can be selected by a line search without evaluating the Hessian matrix. However, the line search method is itself computationally expensive due to the evaluation of multiple error functions. Moreover, the line search deals with multiple parameters crucial for its performance. Sometimes the Hessian matrix becomes non positive definite which leads to increase in the error in weight updates due to ill conditioning in the matrix inversion. In SCG, the Hessian matrix is made positive definite. The computationally inefficient line search in every iteration is avoided using a step size scaling mechanism in SCG. Another advantage of SCG is that it does not involve any user dependent parameters. The SCG method has been used to solve problems in different research domains such as prediction of groundwater level [51], and telemarketing success [52], etc.

Table II reveals that the NMIs between instantaneous frequency and SF (original/regularized) are relatively lower compared to other cases. In the first attempt, two input attributes (seismic impedance and amplitude) have been used to build the prediction model and the corresponding results are documented in Table III in terms of the performance evaluators.

Though variation of instantaneous frequency is comparatively lower, it is an important attribute. Therefore, all three seismic attributes have been used as inputs to the prediction model (Table IV) in the second attempt. The performance comparison of the method of averaging and the proposed regularization procedures has been carried out. The original SF is filtered by a moving average filter with a span of nine samples, before being passed to the ANN training module. Table V represents the validation performance of the ANN in modelling the averaged SF from the seismic impedance, amplitude and instantaneous frequency. The prediction results improve in terms of performance evaluators in case of filtered SF by averaging as compared to the original SF as target. It can be observed from Table IV and Table V that the performance of the trained ANN is superior while working with the processed SF using the proposed regularization as target compared to the original and averaged SF.

The results reported in Table III, Table IV, and Table V lead to the following important observations:
- The performance of the trained networks is improved with the use of regularized target signals. This can be quantified in terms of higher CCs and lower error values.
- In all cases, inclusion of the instantaneous frequency as the third predictor improves the prediction.
- The proposed regularization technique is superior to the averaging method.

Further it can be observed from Table III and Table IV that the network performance is superior in case of regularization based on WD with two predictor variables. On the other hand, with three predictors FT based regularization outperformed the other two regularization approaches in terms of performance evaluators. However, for all cases, the performance is improved while using regularized sand fraction as target instead of original log. The extent of advantage of each scheme over the others may vary among different datasets. Therefore a given scheme cannot be universally recommended. A user is provided with multiple choices among which the user can select the regularization technique that best suits with the working dataset before application of machine learning algorithm.

Literature study reveals that pre-processing is an important step prior to modelling. For example, normalization, relevant attributes selection, importance sampling etc. have been part of the pre-processing technique in different fields such as environmental modelling [53], chemistry [54], biomedical [55], [56], reservoir engineering [17], [57], [58]. Generally, CC values above 80%; RMSE, AEM values below 0.15 and SI value below 0.35 can be considered as good fit in such studies. If the prediction model attains the values of the respective performance indicators beyond the above mentioned limits, then the model should be retrained after adjusting associated parameters. Baziar *et al.* [57] has compared the performance of three machine learning techniques such as SVM, multilayer perceptron (MLP), and Co-Active Neuro-Fuzzy Inference System (CANFIS) to predict permeability from well logs. Further, the normalization has been carried out as the pre-processing technique before modelling. However, integration of seismic and borehole dataset are not required since only well logs are used as the predictor variables in [57]. The performances of the three approaches (SVM, MLP, CANFIS) have been quantified in terms of CC, average absolute error (AAE, equivalent to AEM), and mean squared error (MSE i.e. square of RMSE). The methodology used to model the original SF is almost similar to the methodology adopted for the permeability prediction using multilayer perceptron as in [57]. Comparison between the validation performance (prediction using data not part of training) reported in [57] and those achieved in this study has revealed that the regularization stage improves the prediction result in terms of CC, AEM, and RMSE. The pre-processing step involving a single ANN in [17] is similar to the case in the present paper while modelling with original SF as target. Comparison among the ANN performances with and without the regularization step reveals that the results of ANN has improved in terms of evaluators in the former case.

In the case of unsatisfactory validation performance, the user may modify the ANN structures and network parameters and retrain the modified ANN. If the validation performance still does not improve, the user can modify the regularization parameters and carry out the modelling with regularized target signal. By changing the regularization parameters the information retention in the filtered/reconstructed changes e.g. by choosing a narrower bandwidth in the FT based method the well logs become smoother with a reduced entropy.

TABLE III
STATISTICS OF VALIDATION PERFORMANCE (TWO PREDICTORS)

| Well Name | Original Target Sand Fraction(SF) | | | | Filtered by Fourier | | | | Filtered by EMD | | | | Filtered by Wavelets | | | |
|---|---|---|---|---|---|---|---|---|---|---|---|---|---|---|---|---|
| | CC | RMSE | AEM | SI | CC | RMSE | AEM | SI | CC | RMSE | AEM | SI | CC | RMSE | AEM | SI |
| A | 0.63 | 0.21 | 0.17 | 0.7 | 0.69 | 0.15 | 0.12 | 0.52 | 0.71 | 0.17 | 0.13 | 0.64 | 0.74 | 0.15 | 0.12 | 0.53 |
| B | 0.57 | 0.19 | 0.15 | 0.6 | 0.63 | 0.15 | 0.12 | 0.5 | 0.62 | 0.16 | 0.12 | 0.49 | 0.65 | 0.15 | 0.12 | 0.5 |
| C | 0.68 | 0.16 | 0.12 | 0.53 | 0.78 | 0.12 | 0.09 | 0.37 | 0.76 | 0.12 | 0.08 | 0.37 | 0.77 | 0.11 | 0.08 | 0.36 |
| D | 0.54 | 0.18 | 0.14 | 0.58 | 0.63 | 0.14 | 0.11 | 0.46 | 0.64 | 0.12 | 0.09 | 0.41 | 0.66 | 0.12 | 0.09 | 0.39 |

TABLE IV
STATISTICS OF VALIDATION PERFORMANCE (THREE PREDICTORS)

| Well Name | Original Target Signal | | | | Filtered by Fourier | | | | Filtered by EMD | | | | Filtered by Wavelets | | | |
|---|---|---|---|---|---|---|---|---|---|---|---|---|---|---|---|---|
| | CC | RMSE | AEM | SI | CC | RMSE | AEM | SI | CC | RMSE | AEM | SI | CC | RMSE | AEM | SI |
| A | 0.76 | 0.17 | 0.13 | 0.62 | 0.94 | 0.07 | 0.05 | 0.23 | 0.92 | 0.09 | 0.06 | 0.31 | 0.93 | 0.08 | 0.06 | 0.28 |
| B | 0.65 | 0.17 | 0.13 | 0.56 | 0.86 | 0.09 | 0.07 | 0.30 | 0.88 | 0.09 | 0.06 | 0.28 | 0.87 | 0.10 | 0.08 | 0.34 |
| C | 0.76 | 0.14 | 0.11 | 0.45 | 0.91 | 0.07 | 0.05 | 0.24 | 0.87 | 0.08 | 0.06 | 0.25 | 0.89 | 0.08 | 0.05 | 0.25 |
| D | 0.68 | 0.15 | 0.12 | 0.49 | 0.83 | 0.1 | 0.07 | 0.33 | 0.80 | 0.09 | 0.06 | 0.32 | 0.84 | 0.08 | 0.06 | 0.28 |

TABLE V
STATISTICS OF VALIDATION PERFORMANCE (THREE PREDICTORS, TARGET: AVERAGED SF)

| Well Name | Filtered Sand Fraction (SF) by Averaging | | | |
|---|---|---|---|---|
| | CC | RMSE | AEM | SI |
| A | 0.84 | 0.13 | 0.10 | 0.45 |
| B | 0.77 | 0.14 | 0.10 | 0.44 |
| C | 0.86 | 0.11 | 0.08 | 0.33 |
| D | 0.73 | 0.13 | 0.10 | 0.43 |

The prediction in the entire volume along each inline and cross line has been carried out using the network selected after acceptable performance in the validation step. Fig. 8 represents the variation of the seismic amplitude at a specific inline (inline 136 containing Well D). The sand fraction variation at the same inline shown in Fig. 9 is obtained by prediction of the sand fraction over the study area from available seismic attributes using validated network parameters. The network used here for prediction has been calibrated using EMD–regularized–sand fraction as target.

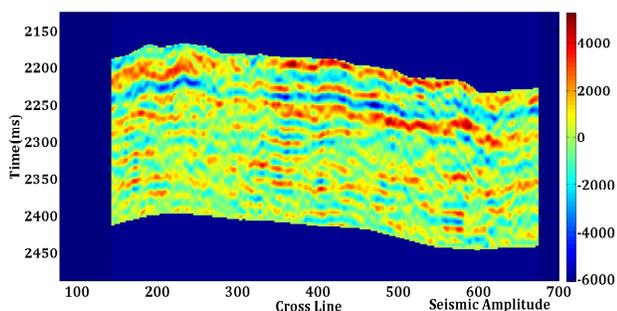

Fig. 8. Variation of the seismic amplitude at inline 136

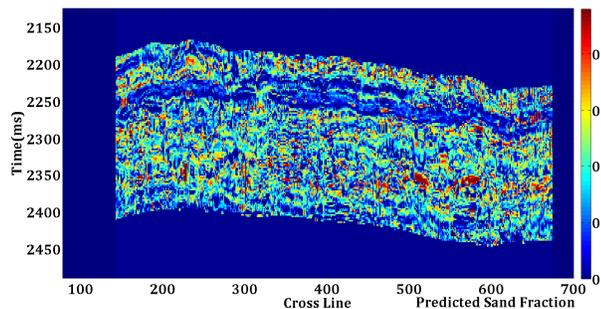

Fig. 9. Variation of the predicted sand fraction at inline 136

## V. POST-PROCESSING

It can be observed from Fig. 8 that the variation of the seismic attributes over the study area is smooth. However, the sand fraction across the study area as shown in Fig. 9 changes abruptly. The transition of the sand fraction values should be smoother and more or less agree to the patterns of seismic data. Thereby rises the need for the post–processing stage in order to obtain a smoother sand fraction variation across the volume. To incorporate this rationale, the predicted values are filtered through a 3-D median filter.

Median filter is an order statistic filter i.e. non-linear spatial filter. In case of order statistics filters, the filtered output is dependent on the ordering (ranking) of the pixels in the image area contained by the selected window [59], [60]. Thus, the neighborhood values of the particular pixel play an important role in the determination of the modified pixel value as a result of order filtering. In case of order statistics filtering the window size is selected to define the neighborhood around the centered pixel. Selection of window value is crucial for the degree of smoothing. The predicted sand fraction in the volume is used as input to the post-processing operation. Every element in the volume is considered as a pixel and is smoothened using 3-D median filter with respect to its neighborhood within a 3x3x3 window size. The missing values along the boundaries are ignored. In order to carry out median filtering, the values of the pixel and its neighbors within the selected window are first sorted. Then, the centered pixel value is replaced by the median value determined from the sorted pixel values. For example, in case of 3-D median filtering with a 3x3x3 window, the 14th largest value of the neighborhood replaces the pixel value at the center of the neighborhood. The SF logs over the area are predicted from seismic attributes and variation of the predicted SF at a particular inline is presented in Fig. 9. The detailing of the SF variation is revealed by the color code in Fig. 9. The smoothened SF values attained by median filtering may lose the exactness of the predicted SF logs; however, the volumetric representation of the filtered SF collectively offers better understanding of the SF variation. Fig. 10 represents the result of median filtering along inline 136. The effect of localizing different levels of sand fraction values can be observed by comparing Fig. 9 and Fig. 10.

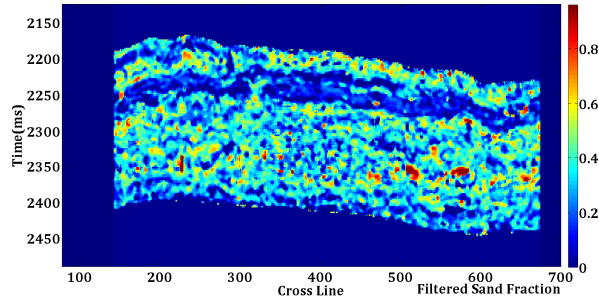

Fig. 10. Result of 3-D median filtering with a 3x3x3 window on the predicted sand fraction variation at inline 136

Post-processing with larger window sizes (for example 5x5x5 or higher order) would further smoothen out the predicted SF volume at the cost of detailed SF variation. In our subsequent research attempts, adaptive post processing method having the potential to further improve the prediction result will be worked out. This would help in subsurface characterization. Different types of spatial filters would be experimented to observe the changes in filtered SF.

Thus, the complete framework including pre-processing, learning and validation, and finally post-processing, successfully carries out mapping between seismic attributes and the sand fraction.

## VI. DISCUSSION AND CONCLUSION

This paper brings out a complete workflow consisting of an elegant regularization step in pre-processing to enhance the learning capability of ANN in carrying out mapping between seismic attributes and lithological property (sand fraction) successfully. The improvement in mapping with introduction of the regularization step is observed from the performance analysis. The comparison among the results achieved in this work and existing literatures [17], [57] reveals the superiority of the proposed regularization step to obtain improved performance in terms of the performance evaluators. In the present study, synthetic SF logs are generated over the study area from available seismic information using the validated network parameters. After obtaining SF volume, post-processing is carried out to improve visualization by means of median filtering. Similarly, other important petrophysical properties such as porosity, permeability can be modelled from seismic attributes following the proposed workflow. The volume-wise prediction results of SF and other petrophysical properties (i.e. porosity, permeability) enables the user to identify zones with high sand content and porosity, and higher probability of hydrocarbon presence. Thus, this study would help to identify potential drilling locations of a new well in a study area.

The present work differs from the work done in [17] in terms of pre-processing stage, division of training-testing dataset, adopted machine learning technique, and post-processing algorithm. In addition the well tops and horizon information are used [17] to carry out zone wise division of the overall dataset and modular ANN is applied to model SF from multiple seismic attributes. The performance attained in [17] would be improved with the use of regularized SF log as target instead of original one.

The selection of initial parameters are crucial for achieving acceptable performance of ANN. In the present work, the ANN structure and activation functions in hidden and output layers have been empirically adjusted. Also the initial values of weight and bias matrices have been chosen randomly. As a possible future extension, the domain of metaheuristic algorithms can be explored in order to automate such selections to further strengthen the framework. Additionally, while in the post-processing stage, the use of spatial filtering has provided a significant improvement in the sand fraction variation over the study area; adaptive post–processing method can be investigated in future.

## APPENDIX
### THE BACK-PROPAGATION LEARNING OF AN ANN USING SCALED CONJUGATE GRADIENT LEARNING

The back-propagation learning of a multilayer perceptron is carried out in two phases. The synoptic weights of the network are constant and the input signal is propagated through the hidden layers to the output layer in the forward phase. The changes only take place in the activation potentials and output of the neurons [27]. In contrast, an error signal is computed as the difference between network output and actual (desired) target (response). The error signal is propagated through layers in backward direction from output layer to input layer in the backward phase. The training procedure is described in details as follows.

Assume,

$$\tau = \{x(n), d(n)\}_{n=1}^{N} \qquad (7)$$

$\tau$ to be the set of training samples used for learning of a back propagation multilayer perceptron. The input vector $x(n)$ is applied to the input layer nodes and the desired output vector $d(n)$ is present at the output node to compute the error between the desired and the actual output. The forward signal is propagated from the input layer to the output layer through single or multiple hidden layers. The induced local field for neuron $j$ in layer $l$ can be computed from the output of neuron $i$ in the previous layer $(l-1)$ at iteration $n$ i.e. $y_i^{l-1}(n)$ and the synaptic weight $w_{ji}^l(n)$ that is connected from neuron $i$ in $(l-1)$ layer and is expressed as

$$v_j^{(l)}(n) = \sum_i w_{ji}^{(l)}(n) y_i^{(l-1)}(n) \qquad (8)$$

In case of output layer (here, $l = L$ and $L$ is the network depth), the output of neuron $j$ is written as

$$y_j^{(L)}(n) = o_j(n) \qquad (9)$$

Therefore, the error is computed as

$$e_j(n) = d_j(n) - o_j(n) \qquad (10)$$

The supervised learning of a multilayer ANN can be viewed as a problem of numerical optimization. The error surface of a multilayer ANN is a nonlinear function of weight vector $w$. Assume that, the error energy averaged over the training samples or the empirical risk be $\xi_{avg}$. Using (10), $\xi_{avg}$ can be computed as

$$\xi_{avg} = \frac{1}{2N} \sum_{n=1}^{N} \sum_{j \in C} e_j^2(n) \qquad (11)$$

where, the set $C$ contains all the neurons in the output layer. The second derivative of the cost function $\xi_{avg}$ with respect to the weight vector $w$ is called the Hessian matrix and denoted by $H$ so that,

$$H = \frac{\partial^2 \xi_{avg}}{\partial w^2} \qquad (12)$$

The Hessian matrix is considered as positive definite unless mentioned otherwise. There are several algorithms to train an ANN. In case of conjugate gradient methods, the computational complexity and memory usage are large because of calculation and storage of the Hessian matrix at each stage. The indefiniteness of $H$ is handled by a scaling coefficient $\lambda_k \geq 0$ in case of the scaled conjugate gradient (SCG). The other parameters $\tilde{r}_k$ and $\tilde{p}_k$ represent the search direction and the steepest descent direction respectively. Different training algorithms have been experimented to train the single layer ANN to model SF from multiple seismic attributes. Finally, the SCG algorithm is selected over other algorithms to train the ANN due to its superiority in terms of obtained performances evaluators. The steps associated with the SCG can be presented as follows [50]:

| **Algorithm IV**: Scaled Conjugate Gradient (SCG) to train a network |
|---|
| **Task**: Obtain weight vector after training |
| **Input**: Training dataset, weight vector, and parameters associated with training |

a) Selection of parameters: Weight vector $\tilde{w}_1$ and $0 < \sigma \leq 10^{-4}$, $0 < \lambda_1 \leq 10^{-4}$, $\tilde{\lambda}_1 = 0$. Assume,
$\tilde{p}_1 = \tilde{r}_1 = -E'(\tilde{w}_1), k = 1$ and success = true

b) If success = true, then compute the second-order information:

$$\sigma_k = \sigma / |\tilde{p}_k|$$
$$\tilde{s}_k = (E'(\tilde{w}_k + \sigma_k \tilde{p}_k) - E'(\tilde{w}_k))/\sigma_k$$
$$\partial_k = \tilde{p}_k^T \tilde{s}_k$$

c) Modify $\partial_k$: $\partial_k = \partial_k + (\lambda_k - \tilde{\lambda}_k)|\tilde{p}_k|^2$

d) If $\partial_k \leq 0$, then make the Hessian matrix positive definite

$$\tilde{\lambda}_k = 2(\lambda_k - \partial_k/|\tilde{p}_k|^2)$$
$$\partial_k = -\partial_k + \lambda_k |\tilde{p}_k|^2$$
$$\lambda_k = \tilde{\lambda}_k$$

e) Evaluate step size: $\mu_k = \tilde{p}_k^T \tilde{r}_k$
$$\alpha_k = \mu_k / \partial_k$$

f) Compute the comparison parameter:
$$\vartriangle_k = 2\partial_k [E(\tilde{w}_k) - E(\tilde{w}_k + \alpha_k \tilde{p}_k)] / \mu_k^2$$

g) If $\vartriangle_k \geq 0$, then error can be reduced:
$$\tilde{w}_{k+1} = \tilde{w}_k + \alpha_k \tilde{p}_k$$
$$\tilde{r}_{k+1} = -E'(\tilde{w}_{k+1})$$
$$\tilde{\lambda}_k = 0, \text{ success = true}$$

If $k \mod N = 0$ then
$$\tilde{p}_{k+1} = \tilde{r}_{k+1}$$
else:
$$\beta_k = (|\tilde{r}_{k+1}|^2 - \tilde{r}_{k+1}^T \tilde{r}_k)/\mu_k$$
$$\tilde{p}_{k+1} = \tilde{r}_{k+1} + \beta_k \tilde{p}_k$$

If $\vartriangle_k \geq 0.75$, modify the scale parameter, $\lambda_k = 0.25\lambda_k$
else $\tilde{\lambda}_k = \lambda_k$, success = false

h) If $\vartriangle_k < 0.25$, increase the scale parameter
$$\lambda_k = \lambda_k + (\partial_k(1-\vartriangle_k)/|\tilde{p}_k|^2)$$

i) Check: If the steepest descent direction $\tilde{r}_k = 0$, then $\tilde{w}_{k+1}$ as the desired minimum. Otherwise, $k = k+1$ and go to b).

**Output**: Trained weight vector $\tilde{w}_{k+1}$

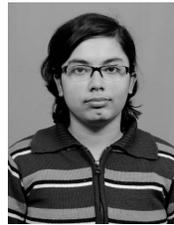

**Soumi Chaki** (S'2014) received the B.E. degree from IIEST Shibpur (formerly BESU, Shibpur), India in 2010. She is currently working toward the M.S. degree in Electrical Engineering from Indian Institute of Technology Kharagpur in India.
   Her research interests include signal processing, machine learning algorithms.

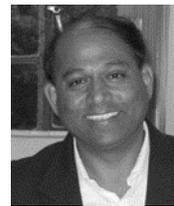

**Aurobinda Routray** (M'2012) is a Professor with the Department of Electrical Engineering, IIT Kharagpur in India.
   His research interests include nonlinear and statistical signal processing, real-time and embedded signal processing, numerical lineal algebra and data-driven diagnostics.

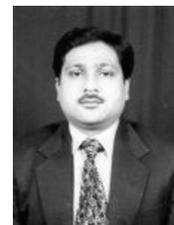

**William K. Mohanty** is a Professor with the Department of Geology and Geophysics, IIT Kharagpur in India.
   His research interests include seismology, reservoir characterization.